# The binaries of the NS-NS merging events

J.E. Horvath[1,a]

[1]*Universidade de São Paulo, Instituto de Astronomia, Geofísica e Ciências Atmosféricas
R. do Matão 1226, Cidade Universitária 05508-090 São Paulo, SP, Brazil*

[a]*Corresponding author: foton@iag.usp.br*

**Abstract.** We discuss in this contribution some issues of the NS-NS merging events, emphasizing that the data and independent evidence gathered from NS mass distribution suggest an asymmetric system as the progenitor of GW170817. The issue of the maximum static NS mass (likely found in redback-black widow systems) is in tension with the value related to this particular event.

## MERGERS AND KILONOVAE: SOME IMPORTANT QUESTIONS

The detection of the GW170817 event, and all the rich emission associated to it in the electromagnetic spectrum, has prompted a great deal of activity trying to understand this and future events. Previous "kilonova" detections were tentatively associated to "NS-NS" mergers (the quotations serve to remind us that the exact nature of these stellar interiors is not known), but it is now proven that these afterglows are indeed associated to a merging, as demonstrated by the gravitational waves/short GRB event. A big bonus is the emergence of multiwavelength data, which is still not understood to the point which the full dynamics and geometry of the event can be consistently explained. However, some of the general features of mergings can be considered now as quite well established. The very emergence of a kilonova has been associated to the presence of lanthanides, required to explain the time behavior of the lightcurves. Spectroscopic data taken after the GW170817 event shows a bump around $1.4 \mu m$ which is probably a signature of a blended bunch of lines from these elements (even though it is unlikely that *individual e*lements can be ever identified in this region). Moreover, it turns out that the geometry of the view is very important, since at least two different components are ejected and contribute to the opacity and hence to the lightcurves as follows.

According to many simulations (i.e. Ref.[1]), the merging of two compact stars produce a cone-like ejection of mass in the direction perpendicular to the orbital plane, which is quite energetic and referred as "dynamical" or "squeezed polar", and on average is not extremely neutron-rich. The second important ejection, along the orbital plane ("secular" or "tidal tail"), has a lower velocity and is more neutron-rich. Because of these features, they have been dubbed the "blue" and "red" component of the ejection. As a warning it is important to stress that it is entirely possible that a third component, the ejection by a wind from the disk, can be the dominant component in mass, perhaps ejecting 10 times as much as the "blue"+"red" components. From the GW170817 data, the first two ejections amount to $\geq 10^{-3} M_\odot$, a value which is in fact larger than most pre-event expectations.

An important issue in this discussion is related not only to the intrinsic geometry of the ejection(s), but also to the relative orientation of the orbital plane and the line of sight. Figure 1 shows a cartoon of the geometry of the event as seen by a distant observer. Three cases are displayed, corresponding to a polar, intermediate and edge-on viewing angle $\theta$. We see that not only the GW emission itself would be quite different, but also the intensity of the blue and red components should be variable, due to the displayed occultation. Fortunately, there is hope to measure the viewing angle for values even greater than the opening angle of the polar emission, provided a sufficient number of events can be collected with the TO time conceded in large optical telescopes [2].

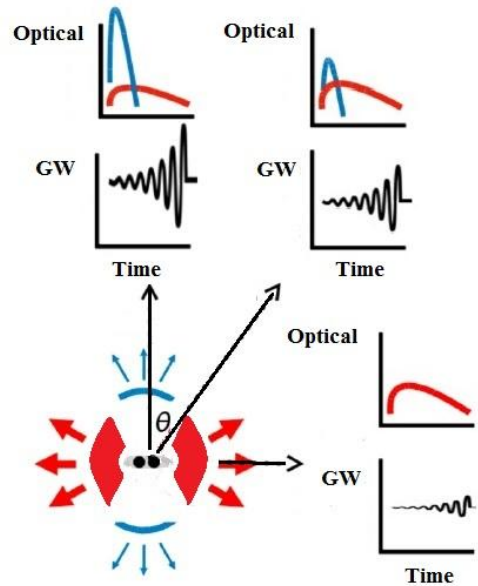

**FIGURE 1.** The issue of the geometry viewing for the determination of the "blue" and "red" components in each event. Three different viewing angles $\theta$ are shown, and the relative intensities of both components sketched, together with the expected gravitational signal (in black).

In addition to these geometrical considerations, the very amount of ejected masses, their neutron fractions and other features would depend not only of the mass ratio of the binary (see below), but also of the *composition* of matter being ejected. The extreme case of this statement is when the "NS" are just exotic objects, made of quark matter. Work is being performed in both the case of "hybrid" configurations [3] and the "strange" stars version [4]. In the latter case, one should consider that the nucleosynthesis yield should proceed from the quark phase (strange quark matter), that fragments into droplets (strangelets) and later decays into ordinary matter. It has been shown [5] that if the fraction of the produced $n/p$ reaches the equilibrium value (as it does in Big Bang nucleoshynthesis), then it will be impossible to produce anything heavier than the iron-peak isotopes, and no lanthanides or "third-peak" r-process elements can be justified in the GW170817 event starting from the hypothesis of a SS-SS merger. However, the dynamics of the expansion and decay channels may preclude chemical equilibrium to be achieved, and the actual production of isotopes in the exotic case is still an open issue under study.

We shall discuss in the next section a related issue, which is independent of the actual composition but quite important for the characterization of the merging population and forthcoming events, namely the mass symmetry of the coalescing systems.

## WHAT KIND OF BINARY PRODUCED GW170817?

As a reasonable working hypothesis, quite consistent with the majority of the known NS-NS systems, most works analyzing GW170817 have assumed a *symmetric* system, in which both stars feature two identical 1.37 $M_\odot$ neutron stars. In spite that most of the measured binary NS systems can be classified as "symmetric", a more careful analysis of both the GW170817 data and the theoretical expectation are in order. Let us start with the observed distribution of NS. A handful of analysis of the "standard sample" [6] are now available [7, 8], indicating that a "single-mass" distribution is no longer tenable. At least two peaks, and more likely three ones have been identified, and a multimodal preference over the single-scale one is now undisputed. For the sake of definiteness, we shall stick to our own results reported in 2011 [8], recently updated with the inclusion of a few relevant additional systems.

Within a Gaussian parametrization (which is a reasonable, but not mandatory choice), the latest numbers indicate that the peaks are located at 1.25, 1.4 and 1.8 $M_\odot$ with standard deviations of 0.07, 0.08 and 0.28 $M_\odot$ respectively, and varying amplitudes. From an evolutionary point of view, it is tempting to track the progenitors which gave origin to this distribution: while the lowest-mass peak is what is expected from the collapse of $O-Mg-Ne$ [9], some low-mass iron cores should be mixed since the lowest measured values of NS (~1.17 $M_\odot$) are actually lower than the minimum former cores. The middle peak is just the "old" value expected for the single-mass hypothesis, and its presence is not very surprising. Finally the highest-mass scale at 1.8 $M_\odot$ contains most of the systems which underwent a substantial amount of mass transfer, unless NS this heavy can be produced directly by very massive iron cores (see below) and do not require accretion.

How does one connect the NS distribution with the problem of the symmetry of the GW170817 event? The waveform analysis has indicated a range of the primary $M_1$ and $M_2$ in the range [1.17, 1.6] $M_\odot$. In addition, the total mass of the system has been measured using the "chirp" mass determination, $M_1 + M_2 = 2.74$ +0.04 -0.01 $M_\odot$. Thus, the probability of the pair as a function of the primary mass subject to the total mass constraint and falling into the pre-determined interval can be constructed extracting the mass values from the reconstructed observed distribution. The result of this simple exercise can be appreciated in Fig. 2. A sketch of the triple-peak distribution (top) which generates the desired $P(M_1,M_2)$ probability (bottom) follows. The issue of "symmetry" is now related to the choice of the mass separation. For definiteness, and taking into account that the mean measured mass in binaries is 1.33 $M_\odot$, we have chosen to consider "symmetric" a system in which $M_1$ lies between [1.33-0.06, 1.33+0.06] $M_\odot$, accounting for the measured dispersion of this subgroup. The result is shown in the lower Fig.2: the probability of the system being outside the dotted lines (and hence being asymmetric) is just above 50%. This is a large probability, not a marginal value, and serves as a warning that an automatic assumption of symmetry may be misleading. This "pos-diction" soundness is further reinforced by the fact that the latest reanalysis [10] of the data favors a mass quotient $q = M_1/M_2 \approx 0.7-0.8$, although it is acknowledged that $q=1$ is not extremely unlikely either.

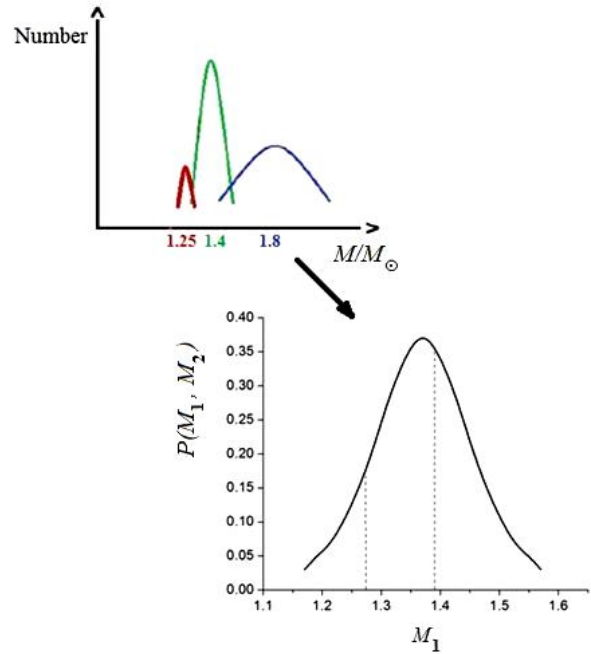

**FIGURE 2.** The construction of a joint probability $P(M_1,M_2)$ (bottom) from the reconstructed NS distribution (top). The dotted lines correspond to $\pm 1\sigma$ around the central average value obtained from binary systems 1.33 $M_\odot$. The area under the curve outside this range is in fact $> 50\%$ by a few percent.

We believe that the *a priori* probability, together with the reanalysis of the data, points towards and asymmetric system merger (see also [11] for a similar conclusion), of the type found by Martínez et al. [12] in the case of PSR J0453+1559. The masses of both objects are reported to be $M_1 = 1.559 \pm 0.005\,M_\odot$ and $M_2 = 1.174 \pm 0.004\,M_\odot$, yielding a value $q \approx 0.75$. However, the measured separation of the of PSR J0453+1559 system is such that it cannot be exactly analogous to the progenitor of GW170817: its merging (coalescence) time is actually larger than the Hubble time $\sim H^{-1}$. Therefore, the real question is whether *sufficiently tight* asymmetric systems can be formed at all. As is well-known, the simplest estimate for the coalescence time is

$$\tau_C = \frac{5}{256} \frac{c^5}{G^3} \frac{a_0^4}{\mu M^4} \tag{1}$$

where $a_0$ is the value of the semi-axis at birth, $\mu$ is the reduced mass of the system and $M$ its total mass. The condition $\tau_C \leq H^{-1}$ demands that the initial separation $a_0$ must be smaller than 2-3 *A.U.*. This is almost impossible to satisfy for the progenitors, with radii comparable to this required separation. In other words, independent quasi-simultaneous supernovae will not be able to produce tight enough systems that merge within a Hubble time. But at least one viable scenario has been recently sketched [13] to give an explanation of an explosive event (iPTF14gqr) identified with an ultra-stripped supernova. The main point of that work is that the supernova happened inside a He-rich envelope, one in which a NS was already present. Therefore, the final state would be a tight double NS system, and the authors claim that this may be the only way to produce very compact NS binaries. Independently of the ultimate truth of this particular evolutionary scenario, it is clear that the nature of the progenitor of GW170817 needs to be explored, even more so if a substantial asymmetry of the masses was indeed present.

## INSIGHTS ON THE MAXIMUM MASS OF NS

The GW signal of the binary NS merger brought another important issue to the consideration of the research in the field, namely the actual value of the maximum mass of a NS in Nature. The connection in this case was achieved by observing the temporal behavior of the GW signal after merger, and also the delay between the GW and the GRB time. According to this picture, either a differentially rotating hypermassive or supramassive configuration was formed, which collapsed to a BH immediately after. Exploiting the well-known relations of rotating configurations in terms of the static parameters yielded a maximum mass of 2.17 $M_\odot$ [14] for the non-rotating, garden variety NSs.

While it has been claimed that this is consistent with the most accurate measurements of massive NSs [14], we would like to point out that these cases may *not* be the highest masses in Nature, and that there is a potentially important tension between those results and actual systems harboring the heaviest NSs produced. The latter group is termed the *black widow-redback* systems, two kinds of interacting close binaries identified in the last years [15]. Work by the La Plata-São Paulo group [16] has worked out the details of the evolution, starting from a restricted region in the donor mass- orbital period for the system to follow an evolutionary track that goes through the redback region (companion masses ~0.1 $M_\odot$, accreting) to the black widow region (companion masses ~0.01 $M_\odot$ or less, being ablated by the NS wind). The most important feature of this novel evolution is that the redback stage is very long, lasting ~ *few Gyr*, and since the efficiency of the accretion rate onto the NS cannot be extremely low, the masses in this stage and in the following black widow group are expected to be very high, independently of their initial values.

A combination of observational spectrophotometric techniques and modeling of the lightcurves and other features has allowed at least three independent groups to report mass values for PSR 1957+20 [17], PSR J1311-3430 [18] and the most recent PSR J2215+5135 [19], namely $2.4 \pm 0.12\,M_\odot$, $2.1-3\,M_\odot$ and $2.27+0.17-0.15\,M_\odot$ respectively. Fig. 3 displays graphically the disagreement between these values and the maximum mass value derived from the GW170817 interpretation. While it may be argued that the error bars and uncertainties are still large, i.e., not as reliable as the Demorest *et al.* pulsar PSR [20] or the Antoniadis *et al.* [21] one PSR J0348+0432, the theoretical calculations reinforce the expectation of very high masses, up to the highest possible value, in these close binary interacting systems. Since the latest calculations of SN explosions do *not* favor the production of NSs to populate the "third peak" bin [22], accretions is left as perhaps the only way to reach the maximum mass value, whatever it turns out to be.

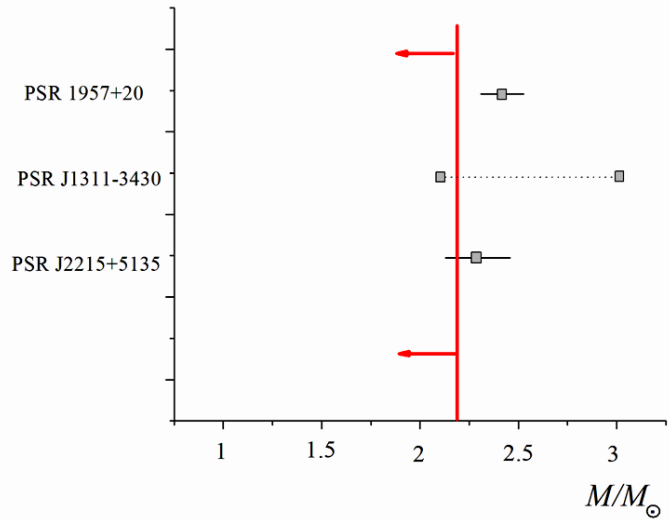

**FIGURE 3.** A comparison between the upper limit derived from the GW170817 signal analysis [14] (vertical line) and the three reported masses for redback-black widow pulsars (square symbols).

These considerations gain a new twist with the recent interpretation that the final remnant of the merger was *not* a BH, but a very massive NS (Dai, these Proceedings, see also [23]). It is not clear whether the afterglow feature claimed in Ref. [23] is real and points towards a re-activation of the engine, as claimed by the authors. But if this is the case, the tension between the upper limit and very massive neutron stars would be gone because the merging value would not apply. In any case, we claim that redback-black widow systems are the true key to the issue of the maximum mass value in actual systems that can be measured with sufficient precision.

## ACKNOWLEDGMENTS

The author wishes to acknowledge the hospitality and financial support of the CUSTIPEN organizers. The Brazilian Agencies FAPESP (through Grant 2013/26258-4) and CNPq 304932/2018-3 have partially financed the present work.